\documentclass{article}
\usepackage[left=2.5cm,right=2.5cm,top=2cm,bottom=2cm]{geometry}
\usepackage[utf8]{inputenc}
\usepackage[T1]{fontenc}
\usepackage{amsmath}
\usepackage{amsfonts}
\usepackage{amssymb}
\usepackage{stmaryrd}
\usepackage{amsthm}
\usepackage{graphicx}
\usepackage{authblk}
\usepackage{caption}
\usepackage{subcaption}
\usepackage{comment}
\graphicspath{images}
\usepackage{soul}
\usepackage{xcolor}
\usepackage[shortlabels]{enumitem}

\begin{document}

\title{Detecting and Correcting Sample-by-Sample Scale Distortion in RNA Sequencing Data}

\author[1]{Christopher Thron}
\author[2]{Farhad Jafari}

\affil[1]{Department of Science and Mathematics, Texas A\&M University-Central Texas, Killeen, TX 76549}
\affil[2]{Department of Radiology, University of Minnesota, Minneapolis, MN 55455}

\date{}

\maketitle


\begin{abstract}

RNA sequencing (RNA-seq) is the conventional genome-scale approach used to capture the expression levels of all detectable genes in a biological sample. This is now regularly used  for population-based studies designed to identify genetic determinants of various diseases. Naturally, the accuracy of these tests should be verified and improved if possible.  In this study, we aimed to detect and correct for expression level-dependent errors which vary from sample to sample, and  are not corrected by conventional normalization techniques . We examined several RNA-seq datasets  from the Cancer Genome Atlas (TCGA), Stand Up 2 Cancer (SU2C), and GTEx databases with various types of preprocessing.  By applying local averaging, we found sample by sample expression-level dependent biases in all datasets studied. Using simulations, we show that these biases corrupt gene-gene correlation estimations and $t$ tests  between subpopulations.  To mitigate these biases, we introduce two  different nonlinear transforms based on statistical considerations that  correct these observed biases. We demonstrate that that these transforms effectively remove the observed per-sample biases, reduce sample-to-sample variance, and improve the characteristics of gene-gene correlation distributions.  Using a novel simulation methodology that creates controlled differences between subpopulations, we show that these transforms reduce variability and increase sensitivity of two population tests. The improvements in sensitivity and specificity were of the order of 3-5\% in most instances after the data was corrected for bias.  Altogether, these results improve our capacity to understand gene-gene relationships, and may lead to novel ways to utilize the information derived from clinical tests.

\end{abstract}

\begin{quote}
\end{quote}

\section{Introduction}\label{sec:intro}

Gathering gene expression data from clinical samples is no longer a technical hurdle, and RNA-seq data is now frequently conducted on biological samples from the clinical and laboratory settings. When considering the holistic profile of gene expression patterns, one can inform of perturbed activity in the cell or tumor models. This information is clinically relevant since dysregulated gene activity is what altogether drives the pathogenesis of the patient's tumors \cite{Qin2020}. The recognition of collective changes in  tumor samples guides pre-clinical research by identifying novel signaling mechanisms and oncogenic processes. Further, this knowledge is purposed towards the development of biomarkers and precision therapies that may extend the survival of cancer patients. Justifiably there is exceptional value in gathering RNA-seq data from clinical specimens; however, quality issues remain due to sequencing platforms themselves or the downstream informatics approaches and data aggregation used to interpret such data. An extensive survey of the sequencing platforms is given in \cite{Survey}. 

In current practice, gene expression levels are estimated based on RNA sequence (RNA-Seq) data obtained from populations of patients. RNA-seq data is generated by isolating RNA from a cell or tissue, converting it to complementary DNA (cDNA), preparing a sequencing library, and using a sequencing platform to convert to transcript counts \cite{kukurba2015rna}. The gene transcript raw counts  for each sample are typically first corrected for fragmentation effects by dividing the raw counts for each transcript by the effective transcript length \cite{kukurba2015rna}. This takes into account the fact that a read can occur at multiple locations along the transcript, so the number of reads per gene will be proportional to the gene transcript length. These corrected counts are then rescaled by multiplying by an overall factor. The two most widely used rescalings are  transcripts per million (TPM) and fragments per kilobase of exon per million (FPKM) \cite{Zhao2021}. In particular, TPM multiplies the corrected counts so that the resulting rescaled counts sum to 1 million.  It is commonly recognized that TPM is more suitable than FPKM for cross-sample comparisons \cite{Zhao2021}. Several other rescalings have also been proposed \cite{robinson2010scaling}. Note that the only difference between these different rescalings is an overall multiplicative factor, which varies from sample to sample.

The gold standard in gene expression analysis is arguably the paper of Wang et al. \cite{Wang2009} (with more than $16,000$ citations) where the authors highlight the strength of RNA sequence data over traditional micro-arrays, particularly in its ability to detect novel transcripts and measure gene expression levels more accurately. They already begin a discussion about biases inherent in such data that is introduced by library preparation steps and sequencing depth. Schurch et al. \cite{Schurch2016}  investigate how different sequencing protocols (e.g. poly-A selection vs. rRNA depletion) can impact gene quantification, and they emphasize the need for standardized protocols. In particular, their conclusions are consistent with the fact that such standardized corrections/metrics do not exist. Robinson et al. \cite{Bioconductor} introduce DESeq method, which addresses biases related to depth and gene length using negative binomial distribution and provide a framework for differential expression analysis while accounting for over-dispersion in count data.
Zhao et al \cite{LiZhao2015} explore GC content bias, sequencing depth bias and the other biases, and present methods to mitigate these biases, including normalization and statistical approaches.  Weiss et al \cite{Weiss2017} use DESeq and recommend sequencing strategies to minimize bias and improve the utility for various of RNA-Seq experiments. Nueda et al. \cite{Nueda2014}, Roberts et al. \cite{Roberts2011}, Soneson et al. \cite{Soneson2016} focus on the challenges of analyzing RNA-Seq data related to batch effects and sequencing platform differences and use various statistical approaches to account for these biases and improve the accuracy of differential gene expression analysis. Yip et al. \cite{Yip2013} review the importance of correcting for biases and propose practical guidelines for addressing them during data processing. This brief summary and the literature cited with their extensive references purport the need for standardized methods to identify and deal with biases that are routinely seen in the sequencing data.

In developing standard normalization methods to deal with biases, the works of (see \cite{McCarthy}, \cite{RO}, \cite{Muren}, \cite{Bioconductor}) can be cited. Specifically the (2021) paper of Feng and Li \cite{Muren} is particularly noteworthy. This paper incorporates the work of Robinson, McCarthy and Smyth \cite{Bioconductor} and a significant amount of literature of the previous decade. Feng and Li base their normalization on a set of
"housekeeping genes" while at the same time questioning the existence of these housekeeping genes and their selection. They carry out pairwise normalization with respect to multiple references, selected from representative samples. Then the pairwise intermediates are integrated based on a linear model that adjusts the reference effects. Motivated by the notion of housekeeping genes and their statistical counterparts, they adopt regression methods and compare their results with existing tools on some standard datasets. Their conclusion suggests that their analysis (MUREN) adjusts the mode of differentiation toward zero, while preserving the skewness due to biological asymmetric differentiation. Their Fig. 1 summarizes workflow. While there is some overlap between their work and ours, their mathematical analysis of the bias and their use of linear regression and housekeeping genes is quite different from our estimation of the bias. Furthermore, we use higher order Taylor expansions to quantify the bias and to construct an inverse function (a transform) to undo the effect of the bias on the gene expression levels. 

Other types of transforms  are used for this purpose \cite{Liu2019}.  The log transform is one widely-used option that has the advantage of being easy to interpret, because 
differences between log-transformed values are roughly proportional to percentage differences in the original data. There are also variance stabilizing transforms (VST)  \cite{love2014differential} which modify the log transform so that different genes have more nearly equal variance, which facilitates comparisons across genes. Each of the many pre-processing steps and normalization approaches has merits and are broadly applied to datasets from consortium efforts as well as clinical reports that include RNA-seq data. 

For research purposes, the  pre-processed data is used to study the tumor biology based on patterns of gene expression. The approaches include conducting differential expression analyses, gene-gene correlations, hierarchical clustering of genes based on expression patterns, enrichment of gene sets, or even machine learning approaches. Researchers adapt these approaches towards identifying the genes that act as drivers or vulnerabilities of the tumors. The recurrently dysregulated genes also have potential as diagnostic tools to evaluate clinical outcomes including overall survival, and  development of metastatic disease  or therapy response.  
Naturally, the methods in which pre-processing is conducted can act as a source of bias when analyzing patterns of gene expression, which reduces the accuracy of our understanding of gene behavior (see \cite{bias}, for example, and the references therein).

This study was initially motivated by preliminary investigations of differential expression levels of genes between two populations of patients within the same dataset. In multiple patient cohorts that exhibited distinct types of cancer, we identified  large shifts in the overall distribution of $t$ statistics when comparing random populations within the cohort. This indicated that there were per-patient biases that shifted the expression levels of all or most genes. We subsequently discovered that these shifts were systematically related to genes' mean expression level \cite{thron2023simple}. If left uncorrected, we hypothesized that these per-patient biases corrupt any biological understanding of patient samples based on conventional forms of downstream analysis.   

 The presence of such anomalous deviations raises several impactful questions:
 \begin{itemize}
     \item Is it possible to identify the cause of the anomalies--specifically, are they due to biological factors, or are there errors in measurement?
     \item Is it possible to correct the effect in order to improve the accuracy of statistical tests?
     \item What effect do the deviations have on statistical tests of RSEM data that are used in medical research?
 \end{itemize}
The goal of this paper is to provide answers to these three questions. 

This paper is structured as follows. In Section~\ref{sec:data} we describe the data used, the transforms used for preprocessing, and some statistical characteristics of the preprocessed data.   In Section~\ref{sec:bias_detect} we identify the bias through block averaging of sorted data. The main contribution of this paper is in Section~\ref{sec:bias_correct}, which describes two alternative approaches to correcting the bias. Section \ref{sec:statTest} will describe the different statistical tests and validations that are performed to demonstrate the effects of corrections to these anomalies. The results are shown in Section \ref{sec:Results} where many statistical studies are performed on the data before and after  the transforms proposed in this article. This is followed by a thorough discussion in Section \ref{sec:Discussion} followed by a summary in the conclusion section. 

\section{The Data}\label{sec:data}

\subsection{Datasets used}
The data used in this study comes from several publicly-available RNA-seq datasets consisting of samples taken from different populations (one sample per patient). For brevity, in the following we will refer to these  samples as "patients".
 The datasets are summarized in Table~\ref{tab:0}. Both raw counts and TPM scaled data were used as indicated in the table. 


\begin{table}
\begin{center}
\begin{tabular}{|l | c | c |} 
 \hline
 Dataset designation & \multicolumn{1}{|p{3cm}|}{\centering Size \\ (patients $\times$ genes)} & Scalings
 \\ [0.5ex] 
 \hline \hline
 TCGA bladder firehose & 
 \begin{tabular}[c]{@{}c@{}}408 $\times$ 20216\\(107 female, 301 male)\end{tabular}
   & 
\begin{tabular}[c]{@{}c@{}}  raw counts,TPM\end{tabular}
 \\
 \hline 
TCGA prostate firehose &
\begin{tabular}[c]{@{}c@{}}498 $\times$ 20198\\(all male)\end{tabular}
 & 
\begin{tabular}[c]{@{}c@{}}  raw counts,TPM\end{tabular}
  \\
 \hline
 SU2C prostate & 
\begin{tabular}[c]{@{}c@{}}208 $\times$ 19158\\(all male)\end{tabular}
& 
\begin{tabular}[c]{@{}c@{}} FPKM, TPM\end{tabular}
\\
\hline
GSE 47774\cite{seqc2014comprehensive} & 
\begin{tabular}[c]{@{}c@{}}1710 $\times$ 39376\\~~\end{tabular}
& 
\begin{tabular}[c]{@{}c@{}} raw counts\end{tabular}
\\
\hline
GTEx prostate V8 & 
\begin{tabular}[c]{@{}c@{}}245 $\times$ 56200\\(all male)\end{tabular}
& 
\begin{tabular}[c]{@{}c@{}} TPM\end{tabular}
\\
\hline
Acute Myeloid Leukemia (TCGA, PanCan) & 
\begin{tabular}[c]{@{}c@{}}173 $\times$ 20531\\~~\end{tabular}
& 
\begin{tabular}[c]{@{}c@{}} raw counts\end{tabular}
\\
\hline
\end{tabular}
\caption{Summary of datasets used in study }\label{tab:0}
\end{center}
\end{table}


All of the tests described in the following were conducted on all datasets. Since results were similar in all cases, the figures are shown for only one case, namely TCGA bladder firehose TPM.

\subsection{Data preprocessing}\label{sec:data_prep}
Datasets with various types of preprocessing were included in the study. The processing used in each dataset is indicated in Table~\ref{tab:0}. In a previous similar study \cite{thron2023simple} we removed outliers of more than 5 standard deviations, but in the current study we did not (there were no noticeable differences between the results).

Regardless of preprocessing, all data was log transformed according to:
\begin{equation}\label{eq:logTrans}
x \rightarrow \log_2(x + c),
\end{equation}
where the constant $c$ is included in \eqref{eq:logTrans} so that unexpressed genes ($x=0$) will not transform to negative infinity. 

Besides the variance-stabilizing property mentioned in the introduction, the log transformation has other advantages. It compresses the range of expression levels from several orders of magnitude to a single order of magnitude, making it easier to compare across genes with widely different expression levels. Most importantly, it transforms multiplicative bias into an additive shift, as follows.

 Suppose that $\{X_j\}$ give the true expression levels for genes  with index $1\le j\le J$. Suppose also that measurement error introduces a multiplicative bias $b\neq 1$, so that the measured values for the genes are $\{bX_j,1\le j\le J\}$. Applying the log transform yields the values 
$\{\log_2(bX_j+c), 1\le j\le J\}$.  If $c$ is small compared to the expression levels $\{bX_j\}$, we may approximate:
\begin{equation}\label{eq:multBias}
    \begin{aligned}
    \log_2(bX_j+c) &= \log_2(b) + \log_2\left(X_j\left(1+\frac{c}{bX_j}\right)\right)\\
    &= \log_2(b) + \log_2(X_j) + \log_2\left(1+\frac{c}{bX_j}\right)\\
    &= \log_2(b) + \log_2(X_j) + \left(\frac{c}{bX_j} + \ldots \right)\\
    &\approx \log_2(b) + \log_2(X_j) + \frac{c}{bX_j}.
    \end{aligned}
\end{equation}
Equation \eqref{eq:multBias} shows that a multiplicative bias in measurement appears as an overall shift across expression levels, if $c$ is small compared to the biased expression level. 
Many studies take $c=1$, but without any explicit justification. We chose  $c=0.25$ for the following reasons.  If measurements are integer values, the smallest nonzero value is 1.  If the true expression levels are rounded to the nearest integer, then values between $0$ and $0.5$ will be rounded to 0. Supposing that small values are approximately uniformly distributed, it follows that the mean of values rounded to 0 is $0.25$. Thus $c=0.25$ is a natural choice.

To minimize the multiplicative distortion introduced by $c\neq 0$, we limited our analysis to genes with mean expression level above 8. For a gene with $X_j=8$, the distortion introduced by $c$ according to \eqref{eq:multBias} is less than $1/32$, or about 3 percent.

In many investigations, the per-gene expression levels is converted to standard scores ($Z$ scores),  so that the per-gene expression levels all have mean 0 and standard deviation 1.  We did not standardize, because the bias' effect on a given gene depends on its mean expression level, but not on its standard deviation. Dividing each gene's expression level by its own standard deviation scrambles the bias so that it can no longer be estimated or corrected. Note however that the statistical tests described in Section~\ref{sec:statTest} do apply standardization to compute $t$ statistics, but only after bias correction.

In the following we will use the term ``expression level'' to denote log transformed expression levels.  We will use the symbols $g$ and $G$ to denote measured and actual values of log transformed expression levels, respectively.

\subsection{Statistical characteristics of preprocessed data}\label{sec:stat_prop}

Many statistical tests depend on the assumption of normality and variance that is independent of expression level. In this section, we verify that log transformed data satisfies these conditions.

\subsubsection{Normality  of expression level sums}
\begin{figure}[ht]
\centering
\includegraphics[width=3.5 in]{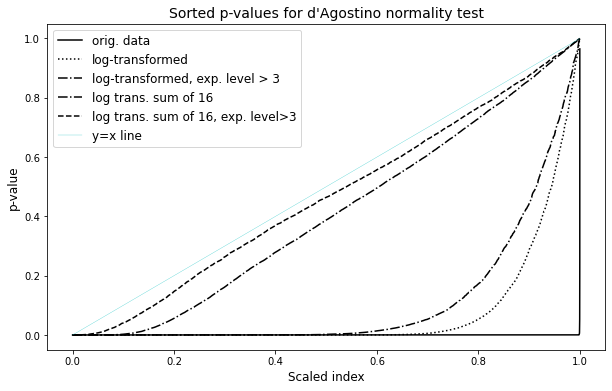}
\caption{Sorted $p$-values for d'Agostino test applied to per-gene data across patients  for untransformed data, log transformed data, and log transformed sums of randomly-chosen samples of size 16. $p$ values restricted to genes with mean expression log transformed expression level above 3 (single genes and random sums of 16 samples) are also shown.  Curves that are closer to the line of slope 1 indicate better normality.}
\label{fig:normal_test}
\end{figure}

In order to characterize the normality of expression level  data, we applied the d'Agostino test for normality to the transformed data for each gene across patients. Figure~\ref{fig:normal_test} shows the sorted d'Agostino $p$-values for all genes for the original data.  A straight line from (0,0) to (1,1) is consistent with perfect normality; however, we find that for the original (untransformed) expression level data, most $p$-values are 0, indicating poor agreement with normality.  
The log transform greatly improves normality, so that about 20 percent of the genes have $p$ values greater than 0.05.  The normality is further improved if only  genes with expression level above 3 are considered (about 44\% of the reference dataset). 

In tests for gene expression level differences between subpopulations, the test statistic depends on per-gene averages across patients in the two subpopulations. In such cases, normality is very closely approximated, even for small patient samples (the improved normality is a natural consequence of the Central Limit Theorem). Figure~\ref{fig:normal_test} also shows d'Agostino $p$-values obtained from the distribution of 16-patient averages from the preprocessed TGCA bladder cancer dataset.   Most of the curve is close to the $y=x$ line, indicating consistency with normality.  Still, about 10\% of genes have $p$-values indistinguishable from 0 on the scale of the figure. But the figure also shows that if  genes with mean transformed expression level below 3 are excluded, the $p$-value distribution is very close to what would be expected from normally-distributed data. These results justifies the assumption of normality in tests that compare mean expression levels for higher-expressed genes between two subpopulations, as long as the subpopulation sizes are greater than 16.

\subsubsection{Uniformity of variance of per-gene expression data}
\begin{figure}[ht]
\centering
\includegraphics[width=5.5 in]{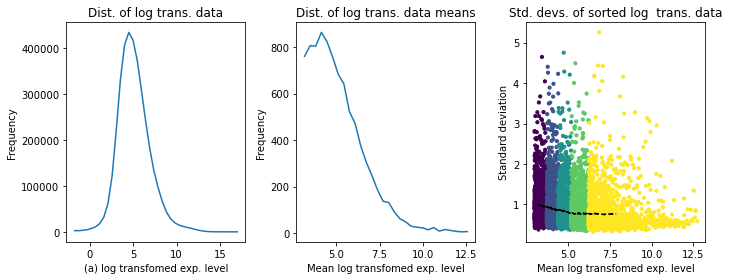}
\caption{(a) Distribution of log transformed expression level for all genes for reference dataset. (b)  Distribution of mean log transformed expression level for genes with mean log transformed expression level above 3.
(c) Standard deviations of log transformed gene expression level versus mean of log transformed expression level, for genes with mean expression level above 3. The color bands indicate quintiles of log expression levels from 0-20\%, 20-40\%, 40-60\%, 60-80\%, and 80-100\%. The black line is the moving average of standard deviations, where each point on the line is the average for 1000 consecutive genes. }

\label{fig:caterpillar}
\end{figure}

We also examine the dependence of per-gene standard deviations on mean expression level for log transformed data.  Figure~\ref{fig:caterpillar} shows the standard deviations of log transformed gene expression levels for the TCGA bladder cancer dataset.  In the plot, genes are ordered by increasing  mean expression level. The black line gives the moving centered block average (block size=800) of gene standard deviations, as a function of gene mean expression level.   The color bands in the plot indicate the ranges for the five data quintiles: for example, about 20\% of transformed expression levels are above 6.

The plot shows that the mean standard deviations of log transformed data are nearly independent of expression level, even without the application of a VST. Furthermore, VSTs in general tend to affect genes with low expression levels. It follows that in studies that focus on mid- to highly-expressed genes, VSTs may provide no advantage, and indeed may introduce unnecessary complications into the analysis, since they do not have the multiplicative shift property that the log transform has.

\section{Methods}\label{sec:Methods}


\subsection{Bias detection and characterization}
\label{sec:bias_detect}
In this section we describe various statistical tests that detect per-patient expression level biases in the studied datasets, and show that they are related to gene expression level.


This task is complicated by the fact that shifts produce a relatively small effect compared to random expression level variation on a per-gene basis. However, the relative effect of the shifts is amplified by taking per-patient block averages of sorted expression level data. 

The procedure for detecting expression-level shifts is described in detail as follows. Note that ``expression level'' refers to the expression level of data that has been cleaned and log transformed as described in Section~\ref{sec:data_prep}.
\begin{enumerate}[(a)]
    \item Compute the per-gene mean expression levels across all patients, and then sort the genes in order of increasing mean value.
    \item Divide the sorted expression levels into $K$ consecutive blocks of equal size, and sum the  expression levels in each block separately for each patient. This step is made mathematically precise as follows. Let $ N, M $ denote the number of genes and patients respectively, and $g_{nm}$ denotes the  measured expression level for gene $n$ $(1\le n\le N)$ and patient $m$ ($1\le m \le M$). Let $B$ denote the block size, and let  $b_{km}$, ($k=1\ldots K, \ m=1 \ldots M$) denote the average expression level for all genes in block $k$ and patient $m$.  Then we have:
\begin{equation}\label{eq:blockAvg0}
\begin{aligned}
b_{km} &= \sum_{j = kL -L + 1}^{kL} g_{jm}, \textrm{~~where~~}  L := \lfloor K/B \rfloor.
\end{aligned}
\end{equation}
\item In order to compare across patients, we then center the values $\{b_{km}\}$ by subtracting out block means across samples to obtain centered block averages $\{\bar{b}_{km}\}$ as follows:
\begin{equation}\label{eq:blockAvg1}
\bar{b}_{km} = b_{km} - \frac{1}{M}\sum_{m=1}^{M} b_{nm}.
\end{equation}
\end{enumerate}
 
\begin{figure}[ht]
\centering
\includegraphics[width=4.5 in]{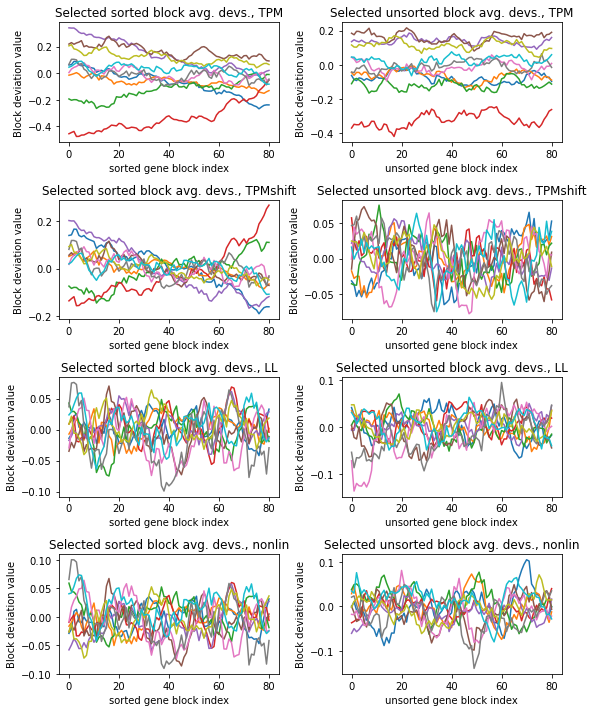}
\caption{Block deviations for selected patients computed according to \eqref{eq:blockAvg1}, for  log transformed TPM data. Each curve gives the block deviations for a randomly selected patient.  Figures in the left column are computed using gene data that is sorted according to increasing expression level, while  the pictures in the right column use the same data but with genes in random order. The top row shows block deviations for log transformed TPM data, using the transform \eqref{eq:logTrans}. The transforms ``TPMshift'' (second row) and ``LLT'' (third row) are explained in Sections~\ref{sec:LLT} and~\ref{sec:NLL}.}

\label{fig:block_deviations}
\end{figure}

Figure~\ref{fig:block_deviations} (\emph{top left}) shows centered block averages as defined by \eqref{eq:blockAvg1} for randomly selected patients from the TPM bladder cancer data, with block sizes of $B= 1000$. The trends show that individual patients have smoothly varying deviations from mean gene expression level depending on expression level.

In order to confirm that the deviations shown in 
Figure~\ref{fig:block_deviations} are dependent on expression level, we repeat the process (1-3) described above, but without sorting the genes by increasing expression level. Figure \ref{fig:block_deviations} (\emph{top right}) shows that block averages of unsorted genes do not show varying deviations. Instead, there are constant overall shifts from patient to patient. This confirms the systematic effect of gene expression level on per-patient expression level deviations from the mean.

A more systematic mathematical analysis of the results is as follows. With $g_{nm}$ as above,
define the average of patient measurements for gene $n$:
\begin{equation}
\bar{g}_n := \frac{1}{M}\sum_{m=1}^M g_{nm},
\end{equation}
and the gene deviation as
\begin{equation}\label{eq:Gmodel2}
\delta_{nm}:= g_{nm} -\bar{g}_n. 
\end{equation}
If we choose a subset of  $N'$ genes $g'_{j_1},\ldots g'_{j_{N'}}$, we have:
\begin{align}\label{eq:Gmodel3}
\frac{\sum_{n'=1}^{N'}g_{j_{n'}m}}{\sum_{n'=1}^{N'}\bar{g}_{j_{n'}}} &=  1 + \frac{ \sum_{n'=1}^{N'}\delta_{j_{n'}m}}{\sum_{n'=1}^{N'}\bar{g}_{j_{n'}}}
\end{align}
For fixed $n$, the values of $\{\delta_{nm'}, 1\le m'\le M\}$ average to 0.  Since the indices $\{j_{n'}\}$ are randomly chosen, it follows that the right-hand side of \eqref{eq:Gmodel3} will be randomly distributed around 0 for different choices of $\{j_{n'},1\le n'\le N'\}$,  unless $\{\delta_{j_{n'}m}, 1\le n' \le N'\}$ is a biased sample from $\{\delta_{j_{n'}m'},  1\le n'\le N', 1\le m'\le M\}$.  The patient-to patient shifts of the curves in the top right panel of Figure~\ref{fig:block_deviations} shows that there is indeed a consistent patient-to-patient bias, regardless of the random choice of  $\{j_{n'}\}$. 
We found similar biases in all datasets studied, regardless of preprocessing. 

The question arises as to whether the shifts observed in per-patient block averages are of biological origin, or if they are due to measurement bias. Two aspects strongly indicate that the cause is measurement bias.  First, the  shifts observed in randomly-selected blocks are consistent, independent of the function of selected genes.  Second, the expression-level dependent shifts vary continuously from block to block across the entire range of expression levels, indicating a global pattern in the genes that once again does not depend on gene function. In view of this conclusive evidence, in the subsequent discussion we will refer to the effect as a bias rather than a shift.

We can easily correct for faulty normalization by changing the multiplicative factor so that the unsorted  block averages for different patients coincide.  The plots in the second row of Figure~\ref{fig:block_deviations} show the results.  The unsorted block averages for different patients now nearly coincide, and average to zero (Figure~\ref{fig:block_deviations}, (second row, right column).  
However, the sorted block averages still show systematic trends (Figure~\ref{fig:block_deviations}, second row, left column). 
Multiplicative normalizations  (such as TPM, DEseq2, and TMM) can only produce uniform vertical shifts in the curves in Figure~\ref{fig:block_deviations}. It is evident that there is some variation between patients which cannot be compensated by an overall normalizing factor. 

\subsection{Bias correction}\label{sec:bias_correct}
In Section~\ref{sec:bias_detect} we demonstrated the existence of per-patient, systematic bias.  This bias can potentially introduce errors into the results of statistical tests,  and may obscure effects due to varying levels of real activity among patients or groups of patients. It behooves us therefore to find a way to  reduce these biases. The most straightforward way to do this is to estimate and then subtract the bias.  The correction problem is then reduced to an estimation problem.

Reference \cite{thron2023simple}  proposes a method termed ``local leveling''  which uses per-patient block averages directly to make corrections to the data. Subsequently it was realized that an approach based on regression can accomplish virtually the same correction but by simpler means.   
In this section we present two alternative regression-based approaches to correcting the patient- and expression level- dependent bias that was demonstrated in Section~\ref{sec:bias_detect}.  The first method (which nearly duplicates local leveling) directly corrects the effect as observed; while the second posits a model for the bias, and derives a corrective transformation based on the model. Thus the first method treats the symptoms of the problem, while the second method attempts to identify the underlying cause in order to provide effective treatment.

\subsubsection{Mean expression level-dependent bias model}\label{sec:LLT}

According to the graphs in Figure~\ref{fig:block_deviations}  the bias in expression level  depends on the genes' mean expression level. A mean expression level-dependent bias gives rise to the following relationship between observed and actual expression levels:

\begin{equation}\label{eq:Gmodel2mean}
g_{nm} = G_{nm} + \beta_m(\bar{g}_n) + \epsilon_{nm},
\end{equation}
where
\begin{itemize}
\item 
$g_{nm}, G_{nm}$ are measured and true values respectively  of the expression level of gene $n$ for patient $m$;
\item
$\beta_m(\cdot)$ represents the mean expression level-dependent bias for patient $m$;  
\item
$\bar{g}_n$ is the average of patient measurements for gene $n$, i.e.   
$\bar{g}_n = \frac{1}{M}\sum_{m=1}^M g_{nm}$;
\item
$\epsilon_{nm}$ is the residual  measurement error in addition to the bias. 
 \end{itemize}
 
In order to derive an estimate of $\beta_m$ for patient $m$, we subtract $\bar{g}_n$ from both sides of  \eqref{eq:Gmodel2mean} to obtain:
\begin{equation}\label{eq:Gmodel3mean}
g_{nm} -\bar{g}_{n} =  \beta_m(\bar{g}_n) +  (G_{nm} - \bar{g}_n) + \epsilon_{nm}.
\end{equation}
The term $(G_{nm} - \bar{g}_n)$ in \eqref{eq:Gmodel3mean} represents deviations of patient $m$'s expression level for gene $n$ from the measured mean.  


The graphs in Figure~\ref{fig:block_deviations} show that the biases $\beta_m$ varies smoothly depending on average expression level.
Accordingly, we assume a polynomial model for $\beta_m$, so that
\begin{equation}\label{eq:polyModel}
g_{nm} - \bar{g}_n=    \left(b_{m0} + b_{m1}\bar{g}_n + \ldots + b_{mL}\bar{g}_n^L\right) +  (G_{nm} - \bar{G}_n) + (\bar{G}_n - \bar{g}_n)+ \epsilon_{nm},
\end{equation}
where we have also added and subtracted $\bar{G}_n$.
In order to estimate the coefficients $b_{m\ell}, \ell = 0\ldots L$, we take block averages.  We introduce the notation: 
\begin{equation}\label{eq:blkAvg}
\langle q_{nm} \rangle_K = \
\frac{1}{K}\sum_{k=n}^{n+K} q_{km},
\end{equation}
for any quantity $q_{nm}$ that depends on both $n$ and $m$.
If the genes are sorted in order of increasing $\bar{g}_n$ and $K$ is sufficiently small, then the function $\beta_m$ will be almost the same for genes $n,\ldots,n+K$, and to a very close approximation we have
\begin{equation}
    \langle \beta_m(\bar{g}_n)\rangle_K = \beta_m(\langle \bar{g}_n\rangle_K). 
\end{equation}
Applying block averaging to \eqref{eq:polyModel}, 
we then have:
\begin{equation}\label{eq:Gmodel2avg}
\langle g_{nm} - \bar{g}_n \rangle_K  = b_{m0} + b_{m1}\langle \bar{g}_n \rangle_K + \ldots +b_{mL} \left(\langle \bar{g}_n \rangle_K\right)^L + \langle G_{nm} - \bar{G}_n\rangle_K + \langle \bar{G}_n - \bar{g}_n\rangle_K + \langle \epsilon_{nm} \rangle_K.
\end{equation}
At this point we invoke the observations described in Section~\ref{sec:bias_detect}, which indicate that random measurement errors (apart from bias) as well as patient-dependent expression level differences average to 0 when block averaged over large numbers of genes with unrelated functions. This implies that the last three terms in \eqref{eq:Gmodel2avg} represent sums of random variables that experience random cancellation.  It is thus reasonable to approximate these terms as normally-distributed, mean zero random variables. In this case, we are justified in using polynomial regression to estimate the coefficients $b_{m\ell}$.

In summary, we estimate the coefficients of $\beta_m$ by first sorting $\bar{g}$, then dividing $g_{nm}$ and $\bar{g}_n$ into $\lfloor N/K\rfloor$ blocks of size $K$, and then applying regression to find estimates $\widehat{b}_{m0},\ldots,\widehat{b}_{mL}$ for the coefficients of $\beta_m$. We use weighted regression with weights based on block averages of $\text{Var}(g_{nm})_{m=1,\ldots M}$, since the variation of gene expression levels between patients is the primary source of error. Once these coefficients are evaluated, we obtain an estimate $\widehat{G}_{nm}^{(LL)}$ for the  true expression level $G_{nm}$ by rearranging \eqref{eq:Gmodel2mean} :
\begin{equation}\label{eq:Gmodel2final}
    \widehat{G}_{nm}^{(LL)} = g_{nm} - (\widehat{b}_{m0} + \widehat{b}_{m1} \bar{g}_n+\ldots + \widehat{b}_{mL}\bar{g_n^L}).
\end{equation}

Figure \ref{fig:block_deviations} (third row) shows that local leveling greatly reduces the deviations for sorted and unsorted gene block averages. It also removes the trends from the sorted gene blocks that were present in the shifted TPM data (row two, column 1)

In the following discussion, we will refer to the above transform as the local leveling transform (LLT). Although it is somewhat different from the local leveling proposed in \cite{thron2023simple}, it leads to substantially the same result.  Both the original and the current LLT make per-patient corrections by averaging over each patient's genes' deviations from the mean value, where the average is performed over genes with nearly the same mean expression level. The original LLT does this averaging explicitly, while the current LLT uses smoothing via regression. 

For completeness, we summarize the statistical assumptions underlying \eqref{eq:Gmodel2final} as follows:
\begin{enumerate}[(A)]
\item 
Measurement bias for a particular gene and patient can be characterized as a polynomial function of the gene's mean expression level across all patients:
\item 
Apart from measurement bias, measurement errors of gene expression levels experience random cancellation when block averaged over large numbers of genes with consecutive expression levels, resulting in a mean-zero random variable that is approximately normal.
\item 
Similarly, differences in a particular patient's actual expression levels from the actual means over all patients also experience random cancellation when block averaged over large numbers of genes with consecutive expression levels, resulting in a mean-zero random variable that is approximately normal.
\end{enumerate}
 
\subsubsection{Nonlinear scale distortion and correction}\label{sec:NLL}

The LL model in the previous section presumes that bias is a function of mean expression level.  This model was motivated by empirical observations described in Section~\ref{sec:bias_detect}.  However, it is more reasonable to suppose that patient-dependent bias in an expression level measurement depends on the expression level in the \emph{patient}, and not the average over all patients. Based on this consideration, we hypothesize that there is a patient-dependent, nonlinear change of scale that distorts gene expression level measurements.  With $g_{nm}$ and $G_{nm}$ defined as above, we may then express the nonlinear relationship between the measurement and true  expression level  as follows:
\begin{equation}\label{eq:znm_model2}
G_{nm} = \alpha_m(g_{nm}) + \epsilon_{nm},
\end{equation}
where
$\alpha_m(g)$ reflects the  expression level-dependent scale distortion for patient $m$, and $\epsilon_{nm}$ represents the measurement error for gene $n$ and patient $m$. 
Assuming that $\alpha_m$ has polynomial form of order $L$ (in  analogy to \eqref{eq:polyModel}), we have:
\begin{equation}\label{eq:alphaPoly}
\alpha_m(g_{mn}) = \sum_{\ell=0}^L a_{m\ell} g_{mn}^\ell
\end{equation}

Taking block averages as in \eqref{eq:Gmodel2avg} and re-expressing $G_{nm}$ in terms of differences as in Section~\ref{sec:LLT}, we have
\begin{equation}\label{eq:blockAvgOfMeas}
\begin{aligned}
    &\langle G_{nm}\rangle_K  = \sum_{\ell=0}^L a_{m\ell} \langle g_{nm}^\ell \rangle_K  + \langle\epsilon_{nm}\rangle_K\\
\implies &\langle G_{nm}-\bar{G}_n\rangle_K + \langle \bar{G}_n-\bar{g}_n\rangle_K + \langle \bar{g}_n \rangle_K = \sum_{\ell=0}^L a_{m\ell} \langle g_{nm}^\ell \rangle_K  + \langle\epsilon_{nm}\rangle_K
\end{aligned}
\end{equation}
Combining the three error terms $\langle G_{nm}-\bar{G}_n\rangle_K, \langle \bar{G}_n-\bar{g}_n\rangle_K$ and $\langle\epsilon_{nm}\rangle_K$ on the right gives a nonlinear model:
\begin{equation}\label{eq:model2Final}
\langle \bar{g}_n  \rangle_K  = \sum_{\ell=0}^L a_{m\ell} \langle  g_{nm}^\ell \rangle_K  + \langle \epsilon^{(tot)}_{nm} \rangle_K,
\end{equation}
where as in Section~\ref{sec:LLT} we approximate $\langle \epsilon^{(tot)}_{nm} \rangle_K$ as mean-zero normal random variables. 
The coefficients $\{a_{m\ell}\}$ can then be estimated by applying regression to \eqref{eq:model2Final}. Because the  error $ \langle \epsilon^{(tot)}_{nm} \rangle_K$  is combined from multiple sources and has no clear relation to gene expression level, we use unweighted regression.  
Then we can estimate expression levels using
\begin{equation}\label{eq:NLmodelfinal}
    \widehat{G}_{nm}^{(NL)} = \sum_{\ell=0}^L \widehat{a}_{m\ell}g_{nm}^\ell,
\end{equation}
which we will refer to hereafter as the nonlinear transform (abbreviated NL).

The assumptions underlying \eqref{eq:NLmodelfinal} parallel assumptions (A)-(C) in Section~\ref{sec:LLT}, except (A) is replaced by:
\begin{enumerate}[($\textrm{A}^\prime$)]
\item 
Measurement bias for a particular gene and patient can be characterized as a polynomial function of the gene's actual expression level for that particular patient.
\end{enumerate}

Figure \ref{fig:block_deviations}
shows that NL has a similar effect on gene block averages as LL. Both remove trends and reduce block deviations equally. 

\begin{figure}[ht]
\centering
\includegraphics[width=4.5 in]{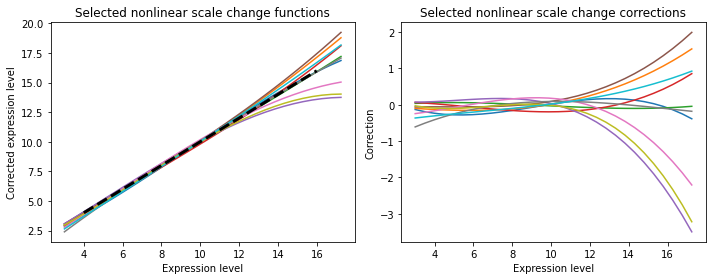}
\caption{Nonlinear scale distortion for a few patients as a function of their gene expression levels. On the left, the measured expression levels are shown as a function of the actual (corrected) levels. On the right, the corrections are shown as a function of the expression levels.}
\label{fig:Nonlinear_scale_chg}
\end{figure}

Figure~\ref{fig:Nonlinear_scale_chg} (\emph{left}) shows the scale change transformation $\alpha_m(g)$ for our reference dataset for five randomly-selected patients, as a function of expression level $z$. The model has polynomial order $L=3$.  Also shown is the $45^\circ$ line, which would correspond to no nonlinear distortion $(\alpha(g)=g)$.  In order to highlight the  scale deviations for different patients, Figure~\ref{fig:Nonlinear_scale_chg} (\emph{right}) shows $\alpha_m(g)-g$ for the same patients plotted in Figure~\ref{fig:Nonlinear_scale_chg} (\emph{left}). The variety of scale deviations highlights the need for separate correction functions for each patient.

\begin{figure}[ht]
\centering
\includegraphics[width=4.5 in]{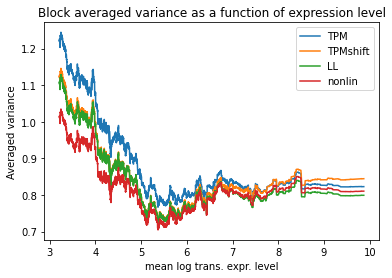}
\caption{Block averaged gene inter-patient expression level variance  for  the different processing methods as a function of mean log transformed expression levels.  }
\label{fig:block_averaged _variance}
\end{figure}

Figure~\ref{fig:block_averaged _variance} displays the block averaged variance for four different transforms of the gene data: original (basic log transformed); shifted; LLT; and NLT. The averaged variance are reduced by NLT for nearly the entire range of the gene expression levels, albeit not much more than LLT. This indicates that NLT is effective in reducing  the between-patient variation.

\subsection{Description of statistical tests}\label{sec:statTest}

We have mentioned that the expression level corrections described in Section~\ref{sec:bias_correct} have the potential of improving the sensitivity and accuracy of statistical tests involving gene expression levels.  In this section we describe several statistical tests that are applied to the data to evaluate the effect of the data transformations described in Section~\ref{sec:bias_correct}. Some of these tests examine statistical properties of transformed data; others involve artificially raising the expression levels of selected genes for selected patients (a procedure that we call ``spiking'') and using standard statistical tests to attempt to identify the spiked genes. These comparisons are carried out both before and after the corrections described here. 

\subsubsection{Correlation tests}\label{subsec:corr}
Expression level data is also commonly used to compute correlations between genes, to determine possible functional dependencies. In order to determine the effect of different transformations on pairwise Spearman correlations, all gene pairs were correlated for all four variant transformations, and the distribution of correlations was plotted.

\subsubsection{Two population tests, single and multiple gene}\label{subsec:twoPopTest}
Genetic expression level data may be used to determine systematic differences between mean expression levels of genes in two populations. For example, an exploratory study might be conducted to determine which genes have significantly higher levels in a subpopulation with particular characteristic. Based on Figure~\ref{fig:normal_test}, we may infer that mean expression levels for sufficiently large subpopulations are nearly normally distributed. Hence the use of a $z$-test or $t$-test is apparently justified.

In order to evaluate the effect of the transformations described in Sections~\ref{sec:bias_detect} and \ref{sec:bias_correct}, we create pairs of subpopulations with controlled differences as follows.  First, we take our actual reference dataset of size $M$ (patients) and randomly separate a subpopulation of size $P$, $ P < M $.  Since the selection is completely random,  any statistical difference detected between these two subpopulations is due to random chance.

In order to introduce a significant difference between the two subpopulations, we add a constant value to the expression levels of genes  the $P$ patients in the selected subpopulation. Since we are dealing with log transformed data, adding a constant $\log_2(C)$ to the data value is essentially equivalent to multiplying the original data  by $C$. We refer to this procedure as ``spiking'', and the value of $C$ is called the ``spike level''. (Note that unlike the "spiking" employed in  \cite{seqc2014comprehensive}, our procedure does not involve modified biological samples, but rather modifications of the data only.)   We then compute $t$ statistics (assuming equal variances for the two populations) for individual genes between the spiked subpopulation  and the remaining population. Note that because of the large number of degrees of freedom, the $t$ statistic is essentially equal to the Z statistic with pooled variance. These $t$ statistics are obtained using the four transforms discussed in Section (log transformed TPM data, shifted log transformed TPM, local leveled, and nonlinear transformed) . The $t$ values thus obtained may be compared to the $t$-distribution obtained between the same two subpopulations, but without spiking.  By so doing, we may determine detection rates of spiked genes. Typically we will be interested in very small $p$ values, because otherwise the test will return too many false positives because of the large number of genes involved.  For example, if the test involves $10,000$ genes, then a $p$ value of 0.005 will return roughly $50$ false positives.

Note that after spiking, patients in the subpopulation are no longer normalized.  However, this does not affect our analysis. Although we are spiking every gene, we compute $t$ statistics gene by gene to model situations where single genes are spiked. We are not performing global comparisons between the spiked and unspiked patients.

The above procedure can be used to conduct two-population tests to detect differences in individual genes' expression levels.  It is also possible to test for statistical significance of differences between multiple genes. In order to conduct such tests on sets of multiple genes, we need to obtain the distribution of sums of gene $t$ statistics.  We may consider genes as a probability space, where each gene has probability weight $1/N$.   Then gene $t$-values define a random variable on this probability space, whose distribution is given by the histogram of gene expression levels.  The distribution of sums of $K$ randomly-chosen  $t-values$, is just the $K$-fold convolution of this distribution with itself.  We may apply $K$-fold convolution to the two subpopulations separately  to obtain the corresponding  $t$ statistic sum distributions.  We may then perform the same analyses on these $t$ statistics that we did on the single-gene $t$ statistics.

\subsubsection{Single-patient tests, single and multiple gene}\label{subsec:spike}
Another possible application of expression level data is the detection of elevated gene levels in specific patients. This may be considered as a special case of the two-population test. Such tests might be used for example to determine whether a certain patient has elevated levels of a cancer-associated gene that has previously been identified. As before, the $t$- or $z$-test may be applied as long as the reference population is sufficiently large.  In this case, the $p$ values of interest are likely to be somewhat larger than in exploratory tests.

It is also possible to test for multiple genes in the single-patient scenario, just as described above for exploratory tests. The more genes are tested for, the more sensitive the test will be to gene elevations of similar magnitude.

\section{Results}\label{sec:Results}

In this section we present all the results that are shown graphically, explaining how they were attained and pointing out the important features in these figures. Further  implications will be given in the discussion section.

\subsection{Correlation tests}\label{results:corrTest}

Figure \ref{fig:Correlation_dist} (\emph{top}) shows the distribution of gene-gene Spearman correlations for the four transforms described in Section~\ref{sec:Methods}. These distributions are compared with the correlation distribution obtained from uncorrelated data generated by independently shuffling each gene's TPM data (thus eliminating any biology-related correlations).
In comparison with uncorrelated data, all four transforms have correlation distributions with a much greater spread, indicating the presence of numerous biology-related correlations. 
TPM  has the largest spread, followed by TPM shifted which has a slightly larger spread than LLT and NLT. TPM also shows a noticeable positive shift, while the other three transforms are nearly symmetric  around zero. 
 This shift is an expected consequence of the expression level-dependent block shifts observed in Figures~\ref{fig:block_deviations}, which introduce artificial correlations.  On the other hand, the correlation distributions for all three transforms are centered at zero, indicating that the net bias introduced by block shifts has been removed.  


Figure~\ref{fig:Correlation_dist} (\emph{bottom}) shows the distributions of differences between correlations calculated using NLT and the other three transforms.  NLT was taken as a baseline, because on theoretical grounds we expect it to give the most accurate correlations.  The figure shows that correlations computed using LLT agree closely with NLT, and TPM shows a broad spread of shifts centered at 0.1.  
The effect of TPM-shift on correlations is much more evident in this plot than in Figure~\ref{fig:Correlation_dist} (\emph{top}): the average magnitude of the difference between TPM-shifted and NLT is roughly 0.05. The broad distribution of differences between TPM-shift and NLT can be attributed to the fact that TPM-shift does not completely remove the expression level-dependent bias, which positively or negatively affects gene pair correlation values depending on whether the pair have similar or dissimilar mean expression levels respectively. Since both LLT and NLT remove the expression-level trends completely, there is no broadening in either transform's distribution of correlations.

The corrective effect of LLT and NLT compared to TPM-shifted is even more evident if we look at correlations of genes with mean expression levels in a narrower range.  Genes with similar expression levels will experience similar per-patient biases, so the overall effect of bias on correlations is amplified. Furthermore, the negative biases that were observed in TPM-shifted data are eliminated in this case, leading to a net positive bias with TPM-shifted data. Figure~\ref{fig:Correlation_dist_limited_range} parallels Figure~\ref{fig:Correlation_dist}
(\emph{bottom}) by showing the distribution of correlation differences, but with only the genes that have mean transformed TPM expression levels  between 4 and 5 (corresponding to a range of 16 to 32 in the original TPM dataset).  The positive skew of TPM-shifted is evident, showing that TPM-shifted does not completely remove effects due to scale nonlinearity.

\subsection{Two population test results}\label{sec:2popTestResults}

Figures \ref{fig:Complementary_50}-\ref{fig:ROC10_multigene} illustrate results of the two-population tests with spiking described in Section~\ref{subsec:spike}. 
All of these figures include results pre-correction (TPM) and post-correction (TPM-shift, LLT, NLT) described in Section~\ref{sec:Methods}.

Figures \ref{fig:Complementary_50}  and \ref{fig:Complementary_10} show mean complementary cumulative distribution functions (ccdfs) for the $t$ statistics obtained using the "spiking" procedure. Recall that the ccdf for a given $t$ level is the proportion of $t$ values which exceed the given $t$ level.  Among other things, ccdfs are used to identify high outliers.   

For population sizes 50 and 10 and spike levels between 20-80 percent, the ccdf was calculated for 5000 randomly-chosen subpopulations. The curves shown are the means of the 5000 ccdf values for each value of $t$. The error bars correspond to the 10th and 90th percentiles  of the 5000 ccdf values obtained from the 5000 simulations.   The error bars for the mean values are too small to be visible on the figure.  

For purposes of comparison,  the same procedure was applied to compute the ccdf for Gaussian data, which is shown in the figure as a dotted black line. The Gaussian data was generated as $J$ independent Gaussian random variables (where $J$ is the number of genes), such that the $j$'th variable has the same mean and standard deviation as the $j$'th gene in the dataset.

We see from the figures for unspiked data that all four transforms  have  mean ccdfs that are very close to the Gaussian ccdf. Error bars for TPM are consistently larger than for the other three transforms, which in turn are much bigger than the Gaussian ccdf error bars.
When spiking is introduced, this situation changes. The mean ccdf's for TPM-shifted, LLT, and NLT are consistently higher than for TPM and Gaussian, with NLT the highest, followed by LLT and TPM-shifted. The difference in error bars between TPM and the other three transforms is more pronounced than for unshifted ccdf's. Comparison between Figures~\ref{fig:Complementary_50} and~\ref{fig:Complementary_10} shows that larger subpopulation sizes yield larger ccdf values, but otherwise produce no qualitative differences.

The high variability in ccdf values justifies the common conclusion that $t$ tests are unsuitable for TPM data, 
even though the mean unenhanced ccdfs  are quite close to the Gaussian ccdf. Nonetheless there is some possibility  of drawing useful conclusions from $t$ tests.  For example, if one applies a $t$ test with $t=3$ to  8,000 genes in a 2-population NLT dataset where the smaller population is 50, we can expect between 0 and 40 false detections (assuming that most of the genes are not enhanced), while genes with enhancement level of at least 40 percent have a 80 percent or higher chance of being correctly identified. If TPM data is used, the number of false detections will be similar, but the minimum percentage of correct positive identifications is somewhat less, at 70 percent.


Figues \ref{fig:ROC_50}-\ref{fig:ROC10_multigene} show various receiver operating characteristic (ROC) curves associated with $t$ tests for two subpopulations. ROC curves are commonly used to evaluate binary classifiers that depend on comparing a statistic to a threshold that is set by the user. ROC curves display sensitivity (also called true positive rate, or TPR) versus 1$-$specificity (a.k.a. false positive rate, or FPR) for the full range of possible thresholds. Since two-population $t$-tests are based on a threshold, ROC curves give an appropriate measure of test performance.  

There are several ways of estimating an ROC curve for a given 2-population test.  
For example, one could use the standard $t$ test ROC curve derived from Gaussian data.  This method has the obvious flaw that  $t$ statistic distributions shown in shown in Figures~\ref{fig:Complementary_50} and \ref{fig:Complementary_10} do not agree particularly well with the Gaussian model. 

A data-based alternative to a Gaussian model is to use resampling. Resampling refers to the procedure of generating empirical probability distributions based on repeated random subpopulation draws from the actual data. There are multiple ways to generate ROC curves via resampling.  One way is to set closely-spaced values of the $t$ statistic $t_1,\ldots t_L$ as thresholds, then use repeated random draws to estimate  the mean TPR and FPR for each threshold.  Another way is to estimate an ROC curve for each random draw; treat each ROC curve as giving TPR as a function of FPR, and average the curves to give the averaged TPR as a function of FPR.  The first method is appropriate when the $t$ threshold is specified before the test---in this case, both the TPR and FPR have uncertainties.  The second method for computing ROC curves is appropriate when the FPR is known, and  the TPR is to be estimated based on this known value---in this case, only the TPR has an uncertainty. In most practical applications, the $t$ statistics are computed, and a fixed number of top-ranked genes is selected as possible true positives. The fixed number provides an estimate for the total detection rate, which approximates the FPR (assuming that most positive detections are false, which is the usual case). Accordingly the empirical ROC curves shown in this paper are calculated using the second method.

Figures \ref{fig:ROC_50}  and \ref{fig:ROC_10} show ROC curves for $t$ statistics obtained using the procedure described in Section~\ref{subsec:spike} on all four transforms, for subpopulations of 50 and 10 patients respectively, with spike levels ranging from 0\% to 80\%. The error bars in the figures give the 10'th and 90'th percentiles for the 5000 draws, as with previous graphs. As before, the error bars on the mean ROC curves themselves are too small to appear in the figure. The differences between transforms are slight, but statistically significant. Of the four transforms, LL and NL tend to have the highest TPR and the smallest uncertainties, while TPM has the lowest TPR and largest uncertainties (except for subpopulation size 50 and spike level 80\%, where these relations are reversed). Comparison between the two figures shows that increasing the subpopulation size greatly increases TPR, especially for lower FPR's.

 The consistent gap between the ROC curves from actual data and from Gaussian data is noteworthy.  A priori there are two possible reasons for this gap: it could be due to the non-normality of the individual gene distributions; or it could be due to correlations between the genes in the actual data.  To determine whether the non-normality is a significant factor, we performed the same simulations, but this time first scrambled the per-gene data among patients. This procedure destroys nonrandom correlations between different genes, but leaves the per-gene distributions unchanged.  When this was done, the ccdf and ROC curves obtained were virtually identical to the Gaussian curves shown in Figures~\ref{fig:Complementary_50}-\ref{fig:ROC_10}, and with similar error bars.  We conclude that the deviations from Gaussian are entirely due to gene-gene correlations in the actual data.

Figure~\ref{fig:detection_by_expression_level}
further breaks down the detectability of outliers as a function of expression level. The figure contains 9 panels arranged as a $3\times 3$ grid, where rows and columns correspond to different enhancement levels and $t$ threshold levels, respectively. Each panel gives $t$ test results from 5000 2-population divisions, where the enhanced population has size 50. 
Genes were ordered by expression level, then divided into blocks of 800, and the average detection rate 
for each block of genes is computed. Each plot shows detection rate versus the mean expression level for all gene blocks, for the four different transforms.  Consistently, we find that genes with lower expression levels also have lower detection rates. This is not surprising given the well-known fact that low-expressed genes have greater variability (see Figure 2 in \cite{thron2023simple}), and are thus harder to detect. 

So far we have only looked at two-population test results for differences in single genes. In Section~\ref{subsec:twoPopTest}, we discussed the possibility of two-population tests for differences in multiple genes, and described a procedure for evaluating distributions of sums of $ t $ statistics by taking convolutions. Figures~\ref{fig:ROC50_multigene} and~\ref{fig:ROC10_multigene} show ROC curves for 2-population multigene tests for subpopulations of size 50 and 10, respectively. Compared to Figures  \ref{fig:ROC_50}-\ref{fig:ROC_10} the differences between transforms are enhanced, especially for the smaller subpopulation.  Comparison across the three panels shows that doubling the number of genes tested roughly doubles the TPR for fixed FPR.

\subsection{Single-patient test results}\label{sec:1PatientTestResults}



Figure~\ref{fig:ROC_1}
shows the averaged ROC curve and 10-90 percentile error bars when the $t$ test was performed for 5000 randomly chosen patients  for enhancement factors of 50, 100, and 200 percent. 
Compared to the larger subpopulation tests, the test is considerably less sensitive (hence the choice of larger enhancement factors in the plots). There is no discernable difference between detection rates for the different transforms.

As with the previous two-population $t$ tests, a gap exists between observed and Gaussian ROC curves. In order to determine the cause of the gap, we scrambled every gene independently among patients and reran the simulation. 
Figure \ref{fig:ROC_1scramble} shows that the ROC curves using scrambled data consistently fell slightly below the Gaussian ROC curves, signifying poorer detection rates. This result distinguishes the single-patient case from the larger subpopulation cases, for which scrambling produced ROC curves identical with Gaussian. We may however draw the same conclusion as before, namely that gene-gene correlations raise the detection rates.

As with the previous 2-population $t$ tests, we also performed multigene tests for the single-patient case. Figure \ref{fig:ROC1_multigene} shows the ROC curve for 5000 simulated single-patient tests for 2, 4, 8 genes at 50 percent enhancement. Not unexpectedly, the TPR for a given FPR doubles as the number of genes in the sample sizes doubles.  There is negligible difference between TPRs for the different transforms, in contrast to  Figures~\ref{fig:ROC50_multigene} and \ref{fig:ROC10_multigene} where NL and LL were clearly superior to unshifted and shifted for multigene tests with multipatient subpopulations.

\section{Discussion and Future Work}\label{sec:Discussion}

Quantitative genomics is a rapidly evolving field that promises to have significant effect in both relating gene expression levels to biological outcomes and to help develop and test drugs that can have important therapeutic effects in treating patients.  However, the complexity of data collection and large uncertainties in quantifying gene expression levels in each patient as well as in populations of patients, natural variabilities between different patients, and many other factors can lead to incorrect intrepretation of the acquired data. Further using the expression levels to establish correlations, function and other vector measures can lead to significantly distorted conclusions. Here we have developed resampling methods that can give us a great deal of valuable information, and improve the veracity and interpretation of statistical tests. 

In this paper, we have demonstrated the existence of nonlinear distortion in RNA-seq data, which depend on expression levels and vary from patient to patient. We have developed two different transformations (LL and NL)  to correct these distortions. We show how these transformations affect important practical statistical estimations. 
The transformations were applied to raw counts, TPM, and FPKM data from several data sets--similarity of the results allowed us to present our results here for TPM data only.  

\begin{figure}[ht]
\centering
\includegraphics[width=4.5 in]{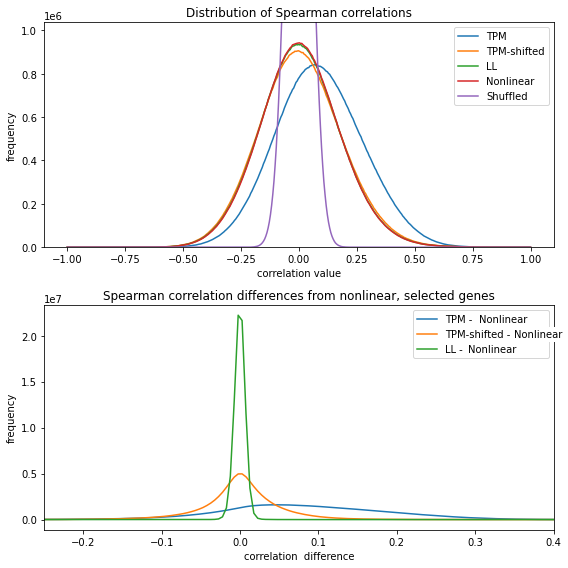}
\caption{(\emph{Top}) Distributions of Spearman correlations for gene-gene pairs subject to four data transforms:  log transformed TPM (abbreviated as TPM), log transformed TPM with shift(abbreviated as TPM-shifted), local leveled (LL), and nonlinear rescaled (NL). The different data transforms are explained in Sections~\ref{sec:bias_detect} and \ref{sec:bias_correct}. For comparative purposes, the correlation distribution for data which shuffles the per-gene expression levels among patients is also shown. (\emph{Bottom}) Distributions of differences between pair-by-pair correlations computed using the different transforms: TPM minus NL, TPM-shifted minus NL, and LL minus NL.  }
\label{fig:Correlation_dist}
\end{figure}

\begin{figure}[ht]
\centering
\includegraphics[width=4.5 in]{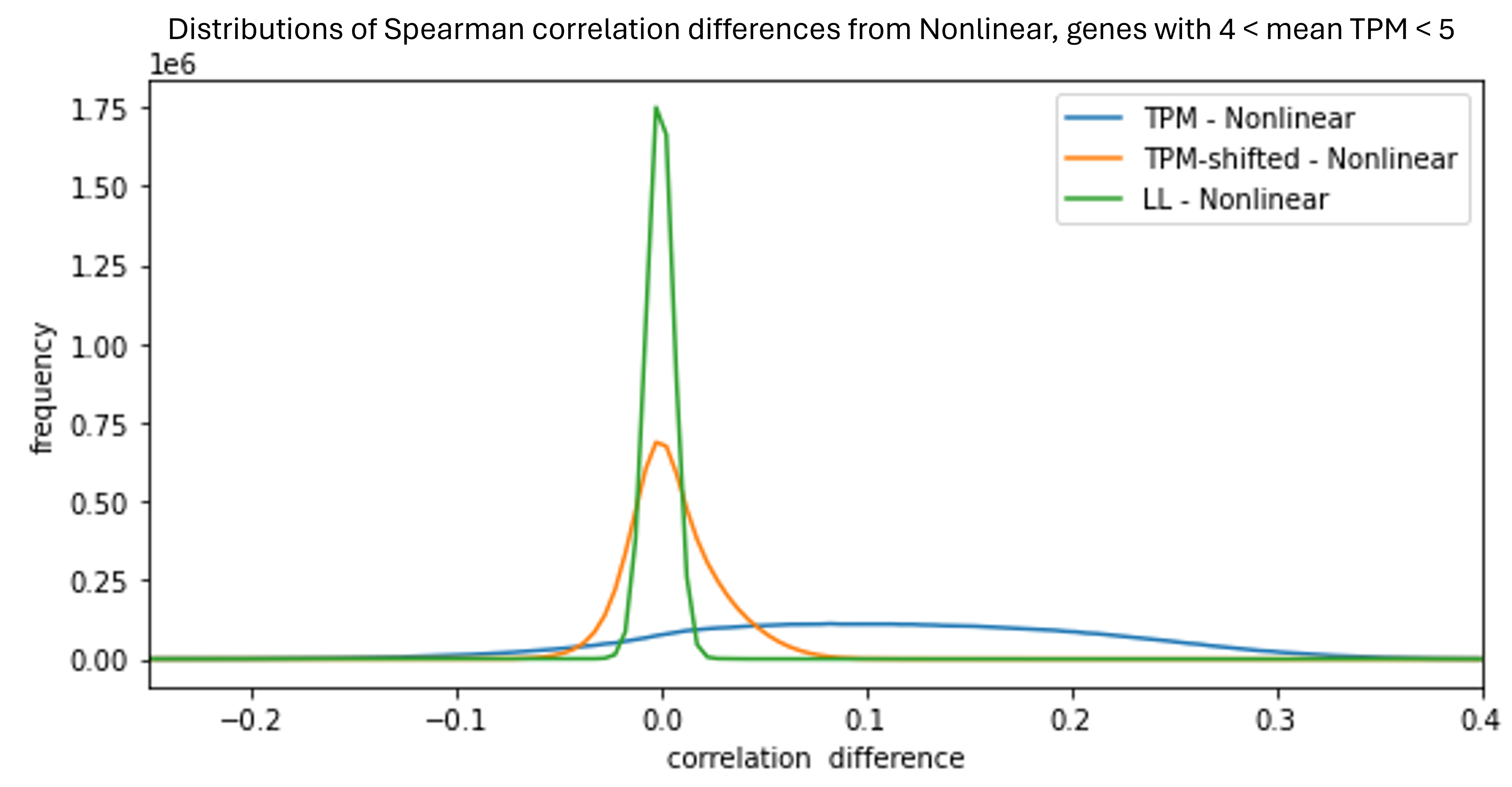}
\caption{Distributions of differences between pair-by-pair correlation differences  as in Figure~\ref{fig:Correlation_dist}(\emph{bottom}), but restricted to genes with mean log transformed expression levels between 4 and 5 (corresponding to TPM values of between 16 and 32). Note the positive skew and overall positive bias in the TPM-shifted correlation estimates.}
\label{fig:Correlation_dist_limited_range}
\end{figure}

\begin{figure}[ht]
\centering
\includegraphics[width=5.5 in]{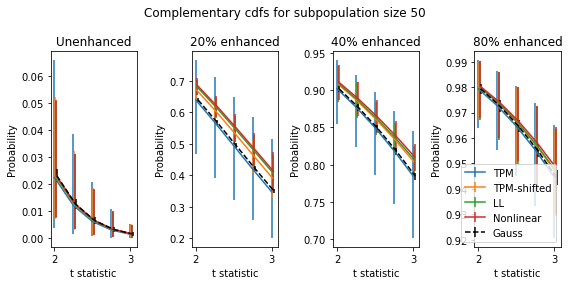}
\caption{Complementary CDFs (ccdfs) for 5000 simulated subpopulations of size 50 at different gene enhancement levels (a.k.a. "spike levels"). For each curve, the $y$-axis values give probability of gene $t$-statistics exceeding the corresponding values on the $x$-axis. Error bars show 10th and 90th percentiles for ccdf values obtained from the 5000 tests: error bars for the averaged curves are too small to see on the figure. The ccdfs for Gaussian data are also shown (see text for full explanation).}
\label{fig:Complementary_50}
\end{figure}

\begin{figure}[ht]
\centering
\includegraphics[width=5.5 in]{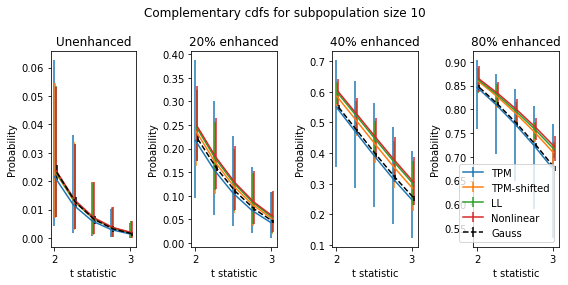}
\caption{Complementary CDFs for 5000 simulated subpopulations of size 10 at different gene enhancement levels. Axes and error bars are as in Figure~\ref{fig:Complementary_50}.}
\label{fig:Complementary_10}
\end{figure}

\begin{figure}[ht]
\centering
\includegraphics[width=6 in]{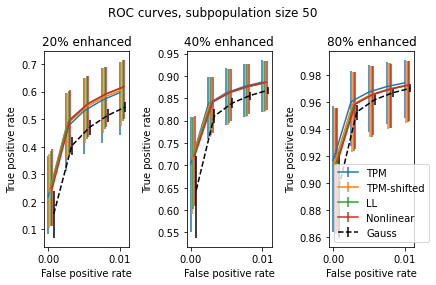}
\caption{Averaged ROC curves for 5000 simulations with randomly-chosen subpopulations of size 50 at different gene enhancement levels (a.k.a. "spike levels"). The $y$-axis is the true positive rate (specificity) for a 2-population $t$, test with false positive rate (1-selectivity) given on the $x$ axis. Error bars show 10th and 90th percentiles. The ROC curve for Gaussian data is also shown (see text for further explanation).}
\label{fig:ROC_50}
\end{figure}

\begin{figure}[ht]
\centering
\includegraphics[width=6 in]{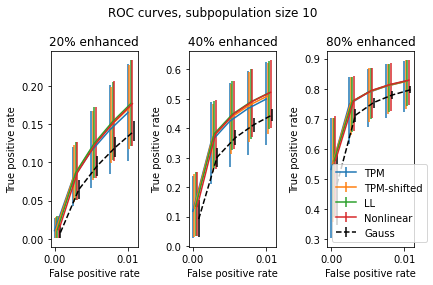}
\caption{Averaged ROC curves for 5000  simulated 2-population tests of size 10 at different gene enhancement levels as shown. Axes and error bars are as in Figure~\ref{fig:ROC_50}.}
\label{fig:ROC_10}
\end{figure}

\begin{figure}[ht]
\centering
\includegraphics[width=6 in]{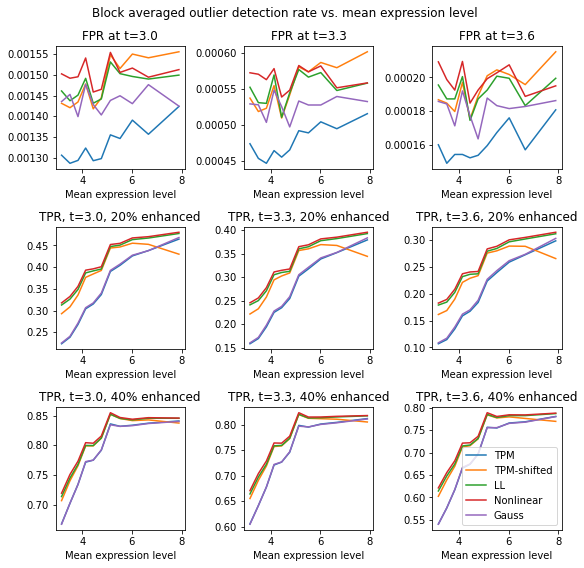}
\caption{Detection rate by mean expression level for gene-enhanced subpopulations of size 50. Rows correspond to different enhancement levels (unenhanced, 20\% enhanced, 40\% enhanced) while columns correspond to different $t$ levels (3.0, 3.3, 3.6). Genes are sorted by mean expression level, and detection rates for blocks of 800 consecutive genes are averaged over 5000 simulations.}
\label{fig:detection_by_expression_level}
\end{figure}

\begin{figure}[ht]
\centering
\includegraphics[width=6 in]{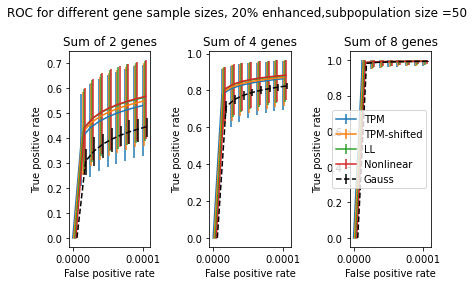}
\caption{Averaged ROC curves for 5000 simulated 2-population $t$ tests with gene-enhanced subpopulation size = 50, at enhancement level 20\%. The three panels show ROC curves for  expression level differences in 2, 4, 8 genes. Axes and error bars are as in Figures~\ref{fig:ROC_50} and \ref{fig:ROC_10}.} 
\label{fig:ROC50_multigene}
\end{figure}

\begin{figure}[ht]
\centering
\includegraphics[width=6 in]{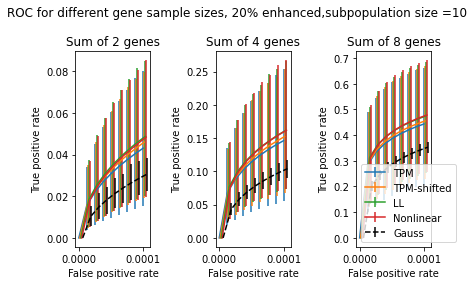}
\caption{ROC curves for  5000 simulated 2-population $t$ tests 2-population $t$ tests with gene-enhanced subpopulation size = 10, at enhancement level 20\%. The three panels show ROC curves for  expression level differences in 2,4,8 genes. (compare Figure~\ref{fig:ROC50_multigene}, which shows similar results for gene-enhanced subpopulation of size 50).}
\label{fig:ROC10_multigene}
\end{figure}

\begin{figure}[ht]
\centering
\includegraphics[width=6 in]{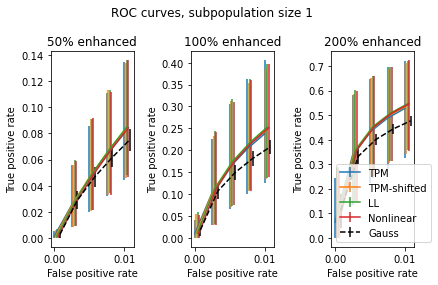}
\caption{Averaged ROC curves for 5000 simulated single-patient  single-gene $t$ tests, for gene enhancement levels of 50, 100, and 200 percent.  Axes and error bars are as in Figures~\ref{fig:ROC_50}-\ref{fig:ROC_10}.}
\label{fig:ROC_1}
\end{figure}

\begin{figure}[ht]
\centering
\includegraphics[width=6 in]{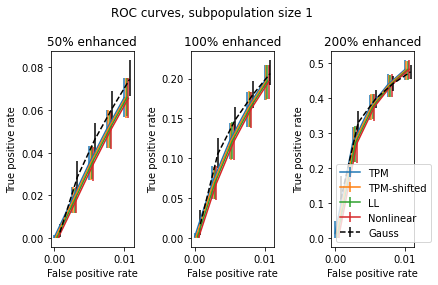}
\caption{Repeat of the test shown in Figure~\ref{fig:ROC_1}, but with each gene's data scrambled separately across patients.}
\label{fig:ROC_1scramble}
\end{figure}

\begin{figure}[ht]
\centering
\includegraphics[width=6 in]{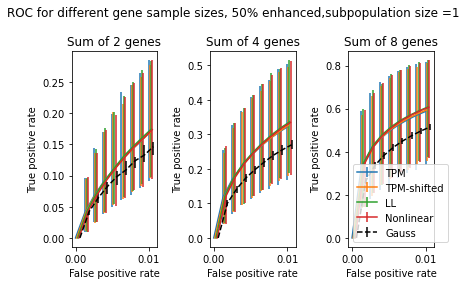}
\caption{ROC curves for  5000 simulated single-patient tests for gene expression level differences, for 2,4,8 genes at 50\% enhancement (compare Figures~\ref{fig:ROC50_multigene} and ~\ref{fig:ROC10_multigene}, which shows similar results for subpopulations of size 50 and 10 respectively).}
\label{fig:ROC1_multigene}
\end{figure}

The analysis can be divided into three parts: (i)  characterize properties of the distortions and the nonlinear corrections as illustrated in figures \ref{fig:block_deviations}-\ref{fig:block_averaged _variance}; (ii) investigate how the distortions affect gene-gene correlations before and after the transforms are applied as illustrated in Figures \ref{fig:Correlation_dist}-\ref{fig:Correlation_dist_limited_range}; (iii) investigate the effect of the distortion on two population test with and without the application of transforms, illustrated by Figures \ref{fig:Complementary_50}-\ref{fig:ROC1_multigene}.  Below we summarize the important points from these three parts.


Figure \ref{fig:block_deviations}  shows evidence of existence of nonlinear scale distortion and the corrective effect of TPM-shifted, LL and NL transforms. The slopes and the shifts of the curves in the top row indicate that distortions vary by expression level and among patients, respectively. The second row shows that TPM-shift can only correct the shift but not the slopes. In fact, any data normalization based on overall multiplicative factors (such as 
Trimmed Mean of M values (TMM), Relative Log Expression (RLE),Transcript Per Million (TPM), or Fragments Per Kilobase
of Mapped reads (FPKM) \cite{corchete2020systematic}) can only shift the curves but not remove the expression level dependent trends. 
Neither can variance-reducing transforms such as VST \cite{hafemeister2019normalization}  or Voom \cite{law2014voom} , since they apply the same transform to all patients so cannot correct for patient-to-patient differences.
However, rows 3-4 show that both LL and NL transforms correct these trends.   

Figure \ref{fig:block_averaged _variance} shows both LL and (especially) NL transforms decrease the block averaged variance of sorted genes,  confirming that LL and NL effectively reduce spurious differences between patients, thereby reducing the inter-patient variance. 
The effectiveness of LL and NL indicates that some recalibration may be required for RNA-seq measurements.


Having established the existence of distortions, we then turn to the consequences of the distortions on the statistical quantities of practical importance.
Figures \ref{fig:Correlation_dist}-\ref{fig:Correlation_dist_limited_range} compare gene-gene correlations before and after LL and NL transformation, revealing a considerable bias in the uncorrected correlation distributions. The positive shift in the distribution computed from uncorrected data indicates that correlations computed with the uncorrected data tend to be overestimates. This suggests that studies that have used correlations based on uncorrected data may need to be re-examined. The symmetry and reduced spread of the LL and NL transformed data is further evidence of distortion corrections. 


Two population tests can be used to identify genetic differences between two subpopulations with different physiological or disease characteristics.  There are many such subpopulation tests that can be devised; we demonstrate the effect the transforms on ccdf, and ROC curves. We developed a spiking methodology to introduce controlled differences between two subpopulations.  $ t $ tests applied to spiked and non-spiked populations can be used to determine test accuracy. Our spiking method can be applied to any dataset without requiring specially prepared samples, as used in \cite{seqc2014comprehensive}. 
The complementary cdf  curves in Figures \ref{fig:Complementary_50}-\ref{fig:Complementary_10} show a nearly $10$\% average improvement in $ t $ test sensitivity due to LL and NL at $20$\% spike level at $t=3$. However the large error bars show that these tests yield highly variable results and are unreliable.

To relate the transforms described to their clinical and diagnostic consequences, we apply the transforms to the gene data and compute average ROC curves for two population $ t $ tests where one population is spiked. 
The ROC data for subpopulations of size $ 10 $ and $ 50$ show about a $3$-$4$\% improvement in TPR at smaller spike levels ($20$-$40$\%) compared to the TPM data, but at larger spike levels ($80$\%) the differences become less noticeable. Since the TPR rates at these higher spike levels are already quite high, the corrections would be more valuable at the early onset of changes in the expression levels. Also, the  TPR error bars for corrected data are smaller than for the original data. As a result,  statistics computed with LL or NL transformed data will be more consistent across studies.  

Detectability of outliers is not uniform across expression levels, and it is more difficult to detect low-expressed genes (cf.  Figure \ref{fig:detection_by_expression_level}). As a result, outliers among low-expressed genes may be under-represented. in comparative studies. It may be possible to compensate for this by applying a variance-reducing transform  (Voom or VST)  subsequent to LL, and NL. Indeed, the two types of transforms should be viewed as complementary, and serve different purposes. NL and LL address inter-patient differences, while Voom and VST compensate for distributional differences between genes.
 
The effects described above that were observed with single-gene, multipatient two population $t$ tests are also seen to varying degrees in multi-gene and single-patient $t$ tests. In general, the improvements in TPR for NL and LL over the baseline TPM  are amplified in multigene and multipatient tests over the single-gene and single patient tests. 



Several areas for future work may be identified. Other RNA-seq datasets may be investigated for the presence of patient-dependent nonlinear scale distortion.  The cause of the distortion may be identified, so that corrections may perhaps be effected at the instrumentation level. The NL and LL transforms may be integrated into an overall pipeline that includes also VST and Voom to provide data on which more sensitive statistical tests may be performed. The resulting pipeline can be utilized to analyze correlations and two-population tests of clinical interest, such as sex, age, race, or disease stage related differences in gene expression.

From a more general perspective, the "spiking" and resampling methods that we have developed may be applied to other tests in statistical bioinformatics to evaluate their accuracy and variability.




\section{Acknowledgements}\label{sec:Acknowledgements}
The authors wish to acknowledge extremely helpful discussions with members of the Hwang Laboratory at the University of Minnesota (Hannah  Bergom, Ella Boytim, Dr. Justin Hwang) and with Dr. Mienie Roberts of Texas A\&M University-Central Texas.

\bibliographystyle{plain}
\bibliography{Bibliography.bib}

\end{document}